\documentclass[11pt,preprint]{aastex}
 
\usepackage{graphicx}
\usepackage{epstopdf}

\slugcomment{Review for the Astronomische Gesellschaft to appear in
Astronomische Nachrichten}
 
\shorttitle{Molecular proto-cluster}
\shortauthors{Carilli et al.}
 
\begin{document}
  
 \title{Radio studies of galaxy formation:
Dense Gas History of the Universe}
 
\author{ 
C.L. Carilli\altaffilmark{1},
  F. Walter\altaffilmark{2},
 D. Riechers\altaffilmark{3},
 R. Wang\altaffilmark{1},
 E. Daddi\altaffilmark{4},
 J. Wagg\altaffilmark{5},
 F. Bertoldi\altaffilmark{6},
 K. Menten\altaffilmark{7}
}

\altaffiltext{$\star$}{The Very Large Array of the National Radio Astronomy
Observatory, is a facility of the National Science Foundation
operated under cooperative agreement by Associated Universities, Inc}

\altaffiltext{1}{National Radio Astronomy Observatory, P.O. Box 0, 
Socorro, NM, USA 87801-0387}
\altaffiltext{2}{Max-Planck Institute for Astronomy, Konigstuhl 17, 69117,
Heidelberg, Germany}
\altaffiltext{3}{Department of Astronomy, California Institute of Technology,
MC 249-17, 1200 East California Boulevard, Pasadena, CA 91125, USA; Hubble 
Fellow}
\altaffiltext{4}{Laboratoire AIM, CEA/DSM - CNRS - University Paris Diderot, 
DAPNIA/Service Astrophysccique, CEA Saclay, Orme des Merisiers, 
91191 Gif-sur-Yvette, France}
\altaffiltext{5}{European Southern Observatory, Casilla 19001, Santiago, Chile}
\altaffiltext{6}{Argelander Institute for Astronomy, University of Bonn, Auf dem H\"ugel 71, 
53121 Bonn, Germany}
\altaffiltext{7}{Max-Planck Institute for Radio Astronomy, Auf dem Hugel 69,53121, Bonn, Germany
}

\begin{abstract}

Deep optical and near-IR surveys have traced the star formation
history of the Universe as a function of environment, stellar mass,
and galaxy activity (AGN and star formation), back to cosmic
reionization and the first galaxies ($z \sim 6$ to 8). While progress
has been truly impressive, optical and near-IR studies of primeval
galaxies are fundamentally limited in two ways: (i) obscuration by
dust can be substantial for rest-frame UV emission, and (ii) near-IR
studies reveal only the stars and ionized gas, thereby missing the
evolution of the cool gas in galaxies, the fuel for star
formation. Line and continuum studies at centimeter through
submillimeter wavelengths address both these issues, by probing deep
into the earliest, most active and dust obscured phases of galaxy
formation, and by revealing the molecular and cool atomic gas.  We
summarize the techniques of radio astronomy to perform these studies,
then review the progress on radio studies of galaxy formation.  The
dominant work over the last decade has focused on massive, luminous
starburst galaxies (submm galaxies and AGN host galaxies). The far
infrared luminosities are $\sim 10^{13}$ L$_\odot$, implying star
formation rates, $\rm SFR \ge 10^3$ M$_\odot$ year$^{-1}$.  Molecular
gas reservoirs are found with masses: M(H$_2$) $> 10^{10}
({\alpha/0.8})$ M$_\odot$. The CO excitation in these luminous systems
is much higher than in low redshift spiral galaxies. Imaging of the
gas distribution and dynamics suggests strongly interacting and
merging galaxies, indicating gravitationally induced, short duration
($\le 10^7$ year) starbursts. These systems correspond to a major star
formation episode in massive galaxies in proto-clusters at
intermediate to high redshift.  Recently, radio observations have
probed the more typical star forming galaxy population ($\rm SFR \sim
10^2$ M$_\odot$ year$^{-1}$), during the peak epoch of Universal star
formation ($z \sim 1.5$ to 2.5). These observations reveal massive gas
reservoirs without hyper-starbursts, and show that active star
formation occurs over a wide range in galaxy stellar mass. The
conditions in this gas are comparable to those found in the Milky Way
disk. A key result is that the peak epoch of star formation in the
Universe also corresponds to an epoch when the baryon content of star
forming galaxies was dominated by molecular gas, not stars. We
consider the possibility of tracing out the dense gas history of the
Universe, and perform initial, admittedly gross, calculations.  We
conclude with a description and status report of the Atacama Large
Millimeter Array, and the Expanded Very Large Array. These telescopes
represent an order of magnitude, or more, improvement over existing
observational capabilities from 1 GHz to 1 THz, promising to
revolutionize our understanding of galaxy formation.

\end{abstract}
 

\section{Introduction}

\subsection{The optical view of galaxy formation}

A dramatic advance in the study of galaxy formation over the last
decade has been the delineation of the cosmic star formation rate
density (the 'star formation history of the Universe'; SFHU) to a
look-back time within 0.6 Gyr of the Big Bang (Madau et al. 1996;
Bouwens et al. 2010). Three main epochs have been identified (Figure
1): the first is a gradual rise during cosmic reionization at $z \sim
10$ to 6, corresponding to the epoch when light from the first
galaxies and quasars reionize the neutral IGM that pervaded the
Universe (Fan et al. 2006). Second is the 'epoch of galaxy assembly'
at $z \sim 1$ to 3, when the cosmic star formation rate density peaks,
and during which about half the stars in the present day Universe form
(Marchesini et al. 2009). And third is the order of magnitude decline
in the comoving cosmic star formation rate density from $z \sim 1$ to
the present, constituting the inexorable demise of galaxy formation
with cosmic time, as we run out of cold gas.

These studies have progressed to the next level of detail, namely the
SFHU as a function of galaxy environment, stellar mass, and star
formation rate (SFR). One interesting result is the observation of
'downsizing' in galaxy formation. This entails the systematic decrease
in specific star formation rate (SSFR; star formation rate per unit
stellar mass), with increasing stellar mass (Cowie et
al. 1997). Downsizing is a manifestation of the general fact that
massive galaxies form most of their stars early and quickly, and the
more massive, the earlier and quicker. Evidence includes studies of
the evolution of the SSFR (Moresco et al. 2010), stellar population
synthesis studies of nearby ellipticals (Renzini 2006; Collins et
al. 2009), and the direct observation of evolved, passive galaxies at
high $z$ (Kurk et al. 2009; Doherty et al. 2009; Andreon \&
Huertas-Company 2011).

A second interesting result is the shift of the balance of star
formation to more actively star forming galaxies with increasing
redshift. At $z \sim 0$, the cosmic star formation rate density is
dominated by galaxies with star formation rates $\le 10$ M$_\odot$
year$^{-1}$ (FIR luminosities $\le 10^{11}$ L$_\odot$). By $z \sim 2$,
the dominant contribution shifts to galaxies forming stars at $\ge 100$
M$_\odot$ year$^{-1}$ (Murphy et al. 2011; Magnelli et al. 2011).

\begin{figure}
  \includegraphics[height=55mm]{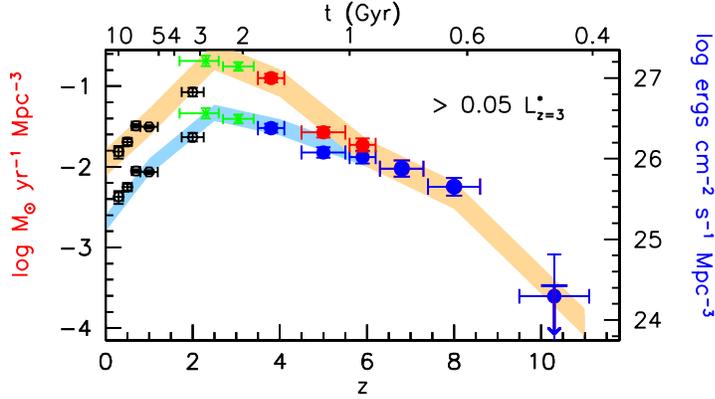}
  \caption{The evolution of the comoving cosmic star formation rate density
as a function of redshift. The blue curve indicates star formation rates
estimated from rest-frame UV measurements with no dust correction. The
yellow curve includes the subtantial dust correction (from Bouwens et al. 
2010).}
\end{figure}

\subsection{The role of radio observations}

The results above are based, for the most part, on observations at
optical through near-IR wavelengths. While truly remarkable in scope,
optical through near-IR studies of galaxy formation are limited in two
fundamental ways. First, dust obscuration plays a substantial role in
determining views of early galaxies, in particular, the most active
star forming galaxies. The average correction factor for
Lyman Break galaxies (LBGs) when deriving total star formation rates
from UV luminosity entails a factor five increase from observed to
intrinsic star formation rates. This technique has been refined
through the use of UV spectral slopes (Calzetti et al. 1994; Daddi et
al.  2004).

\begin{figure}
  \includegraphics[height=75mm]{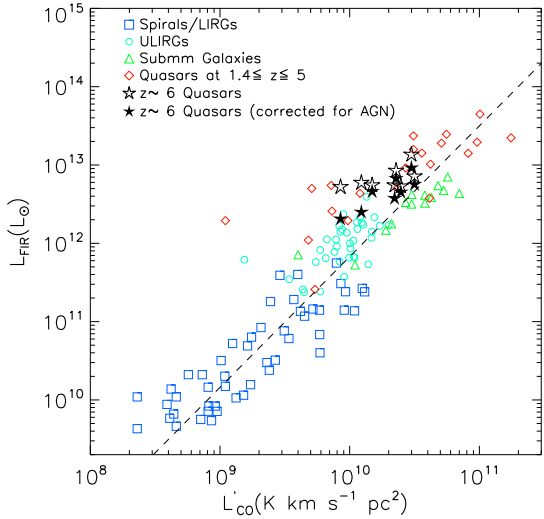}
  \caption{The correlation between Far-IR luminosity ($L_{FIR}$) and CO 
line luminosity ($L'_{CO1-0}$) for both low and high redshift star forming
galaxies (from Wang et al. 2010). 
}
\end{figure}

And second, optical/near-IR studies reveal the stars, but miss the
cold molecular gas, the fuel for star formation in galaxies. There is
a well established correlation between the Far-IR and CO luminosity of
galaxies (Figure 2). This correlation is not surprising, given that
stars form in molecular clouds (Bigiel et al. 2011), and it argues 
that the rapid evolution of star formation with redshift
should be reflected in the evolution of the molecular gas content of
galaxies.

Observations at centimeter through submm wavelengths solve both these
issues. Radio observations probe deep into the earliest, dusty, most
active phases of star formation in galaxies. Similarly, cm through
submm observations reveal the cool molecular and atomic gas in
galaxies. 

\begin{figure}
  \includegraphics[height=100mm,angle=-90]{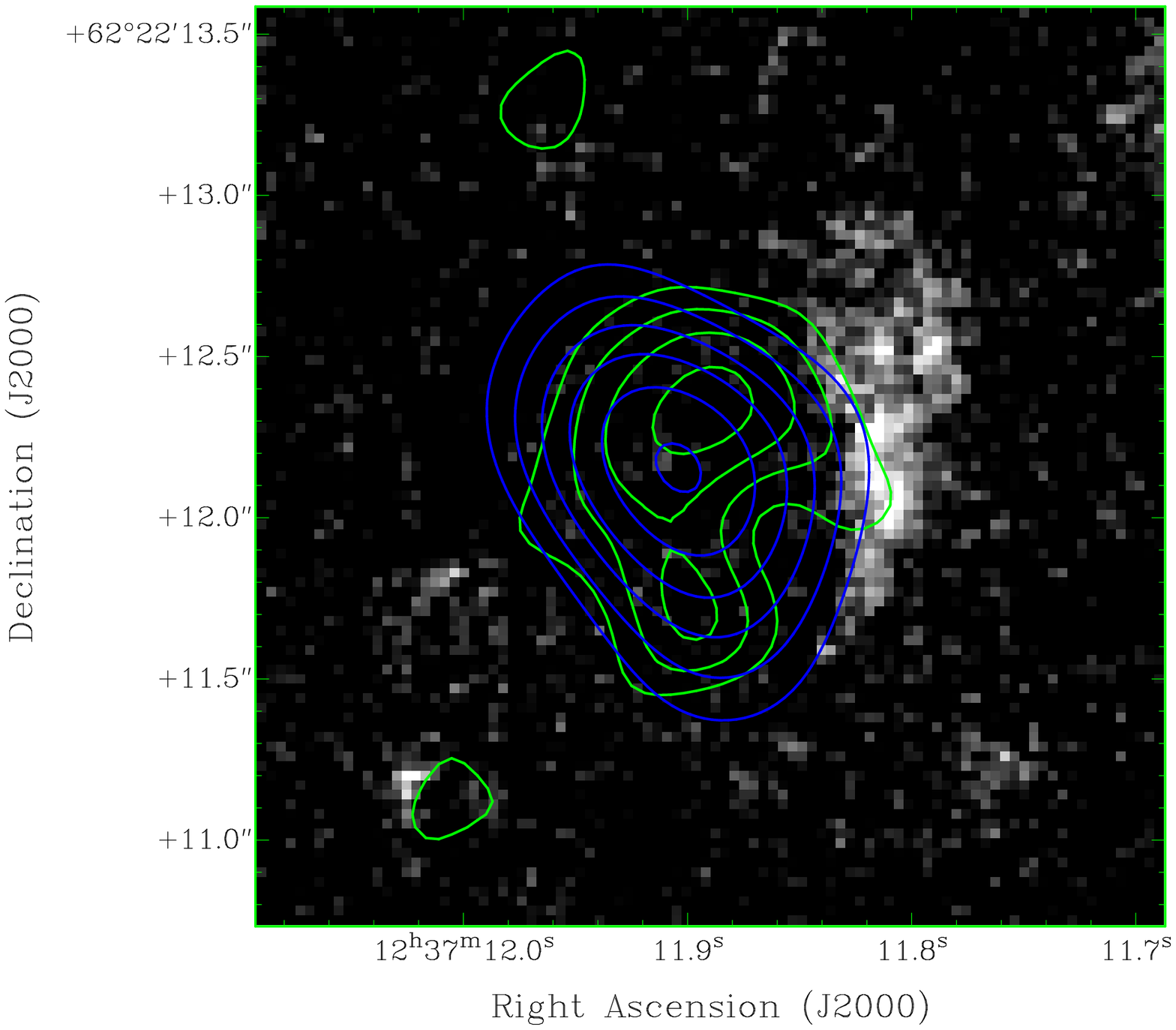}
  \caption{Blue contours show the 250GHz continuum emission from the $z =4.0$
SMG GN20. Green contours show the CO 2-1 emissio. The greyscale shows the 
HST i-band image (from Carilli et al. 2010). 
}
\end{figure}

Figure 3 shows an example of these phenomena in a luminous starburst
galaxy at $z \sim 4$.  The contours show the thermal emission from
warm dust and the CO emission, while the greyscale is the HST
i-band image.  The dust and gas trace the regions of most active star
formation, and these regions are completely obscured in the HST image.

\section{Tools of radio astronomy}

We briefly review some of the observational tools available to study
the gas, dust, and star formation in distant galaxies. 

\subsection{Continuum}

Figure 4 shows the SED at cm through FIR wavelengths of an active star
forming galaxy redshifted to $z = 5$.  The cm emission is synchrotron
radiation from cosmic ray electrons spiraling in interstellar magnetic
fields. These electrons are accelerated in supernova remnant shocks,
and hence the radio luminosity will be proportional to the massive
star formation rate. The FIR emission is from dust heated by the
interstellar radiation field, which in active star forming galaxies is
dominated by massive stars.

\begin{figure}
  \includegraphics[height=90mm,angle=-90]{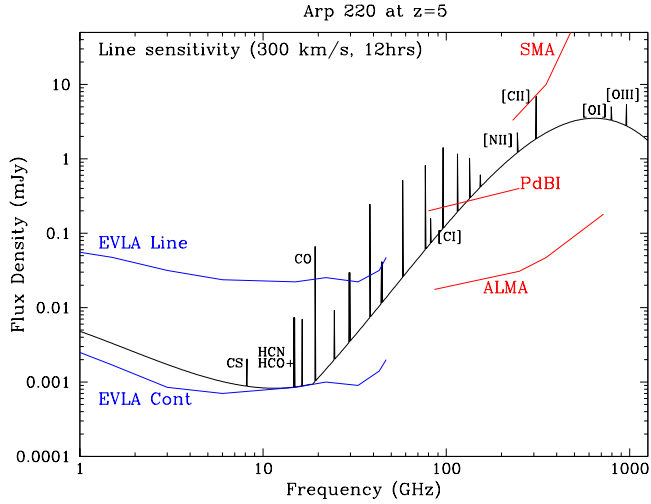}
  \caption{Radio through submm spectrum of a star forming galaxy with a 
star formation rate 100 M$_\odot$ year  redshifted to $z = 5$. Also shown 
are the line and continuum sensitivity for the EVLA in 12 hours, and the line
sensitivity for ALMA and the existing (sub)mm interferometers. 
}
\end{figure}

Hence, both the FIR and radio emission are a function of the massive
star formation rate, resulting in the well quantified correlation
between the radio and FIR luminosity from galaxies (Yun et al. 2001).
While the tightness and linearity of the relationship over such a
large range in luminosity remains puzzling, this correlation is an
important tool for studying dust obscured star formation in galaxies,
providing two means for estimating massive star formation
rates in distant galaxies.  Studies of galaxies at least out to $z
\sim 2$ show that the correlation remains unchanged (Bourne et al.
2011), although at the highest redshifts it may be inevitable that
inverse Compton scattering losses off the CMB by the relativistic
electrons leads to a departure from the low $z$ relationship 
(Carilli et al. 2008).

An important point is that FIR emission peaks around 100$\mu$m. For
high redshift galaxies, this emission shifts into the submm bands.  A
related point is the well studied 'inverse-K' correction in the submm:
the rapid rise in luminosity density on the Rayleigh-Jeans side of the
modified grey body emission curve offsets distance losses, leading to
a roughly constant observed flux density with redshift for a fixed
observing frequency and intrinsic luminosity (Blain et al 2002). In
essence, submm observations provide a distance independent method with
which to study star forming galaxies.

For reference, for a typical dust temperature of $\sim 45$K, the
relationships between observed 350 GHz flux density ($S_{350 GHz}$, in
mJy), dust mass, integrated far-IR luminosity, and total star
formation rate (SFR; Chabrier IMF) are approximately:

$$ L_{IR} = 1.5\times 10^{12} S_{350 \rm GHz} \rm ~L_\odot $$ 

$$ M_{dust} = 7.6\times 10^{-5}~ L_{IR} \rm ~ M_\odot$$

$$ {\rm SFR} = 1.0 \times 10^{-10} L_{IR} \rm ~ M_\odot ~ year^{-1}$$

\subsection{Molecular rotational transitions}

A rich spectrum of rotational transitions of common molecules
redshifts into the cm and mm bands for distant galaxies (Figure
4). Most prominent are the emission lines from CO.  CO has long been
used as a tracer for the total molecular gas mass (dominated by
H$_2$): 

$$\rm M(H_2) = \alpha {\it L'_{CO1-0}}\rm ~~ M_\odot$$

\noindent where $\alpha$ is the CO luminosity to H$_2$ mass conversion
factor. The units for $L'_{CO1-0}$ are K km s$^{-1}$ pc$^2$. These
units were originally designed for spatially resolving observations of
CO in the Galaxy, where brightness temperature was paramount.

Values of the CO luminosity to H$_2$ mass conversion factor $\alpha$,
range from 0.8 M$_\odot$/[K km s$^{-1}$ pc$^2$] for luminous starburst
nuclei in nearby galaxies, to 3.6 M$_\odot/$[K km s$^{-1}$ pc$^2$] for
Milky Way-type spiral disks. Since the CO emission is optically thick
on small scales (ie. molecular cloud cores), $\alpha$ is calibrated
essentially based on dynamical measures of masses, from virialized
GMCs to gas dominated rotating disks in starburst nuclei (Downes \&
Solomon 1998).  There is evidence for both values in the different
populations of high $z$ galaxies, again separated according to compact
starbursts and star forming disk galaxies (Daddi et al 2010b; Tacconi
et al. 2010; Genzel et al. 2010; Naraynan et al. 2011).

Solomon \& vanden Bout (2005) derive the following relationships
between CO luminosity and observed flux density and line width:

$$ L'_{CO} = 3.3 \times 10^{13} S \Delta V D_L^2
\nu_o^{-2}(1+z)^{-3} \rm \ ~~~ K~ km~ s^{-1}~ pc^2 $$

\noindent where $\nu_o$ is the observing frequency in GHz, the
luminosity distance, $D_L$, is in Gpc, the flux density, $S$, is in
Jy, and velocity width, $\Delta V$, is in km s$^{-1}$. For CO
luminosity in solar units the relationship is:

$$ L_{CO} = 1.0\times 10^{3} S \Delta V (1+z)^{-1}\nu_r  D_L^2 ~~~ \rm L_\odot $$

\noindent where $\nu_r$ is the rest frequency in GHz.  Solving for
$S\Delta V$ in (4) and (5), and equating (Carilli 2011), yields:

$$ L_{CO} = 3\times 10^{-11} \nu_r^3 L'_{CO} ~~~ \rm L_\odot $$

Other critical contributions of molecular line observations include:

\begin{itemize}

\item Gas velocities determine the dynamical masses of
high$z$ galaxies, and internal gas dynamics in star forming regions.

\item Multi-transition studies provide the gas excitation, which gives
a rough estimate of gas density and/or temperature.

\item Observations of high dipole moment molecules, such as HCN and HCO+,
  provide an estimate of the dense gas content of galaxies ($\rm n(H_2) > 
10^4$   cm$^{-3}$).

\end{itemize}

\subsection{Atomic fine structure transitions}

The atomic fine structure lines are emitted predominantly in the rest
frame FIR, and hence redshift into the submm range for distant
galaxies (Figure 4). Being metastable transitions, and hence typically
optically thin, these lines, and in particular the [CII] 158$\mu$m
line, are the principle coolant of interstellar gas (Spitzer 1998).
The [CII] can carry up to 1\% of the total IR luminosity from
galaxies, and is typically the brightest line from IR through meter
wavelenths (Malhotra et al. 2001; Bennett et al. 1994). The [CII] line
traces the CNM and photon-dominated regions associated with star
formation (Cormier et al. 2010). Fine structure line ratios can be used
as an AGN versus star formation diagnostic (Genzel \& Cesarsky 2000).

Herschel is providing a revolutionary view of these lines in nearby
galaxies (Cormier et al. 2010). Submm telescopes are beginning to make
serious in-roads into the study of these lines in distant galaxies
(Stacey et al. 2010).

\section{Molecular gas at high redshift}

Figure 5 shows a histogram of the number of detections of CO emission
from high redshift galaxies versus year. To date, there are
115 detections of CO at $z > 1$. The number of detections
has almost doubled in the last two years  due to two factors. 
First is improved instrumentation, in particular continued improvements
at the Plateau de Bure Interferometer (PdBI), and the coming on-line
of the EVLA and the GBT Zpectrometer.  

\begin{figure}
  \includegraphics[height=60mm]{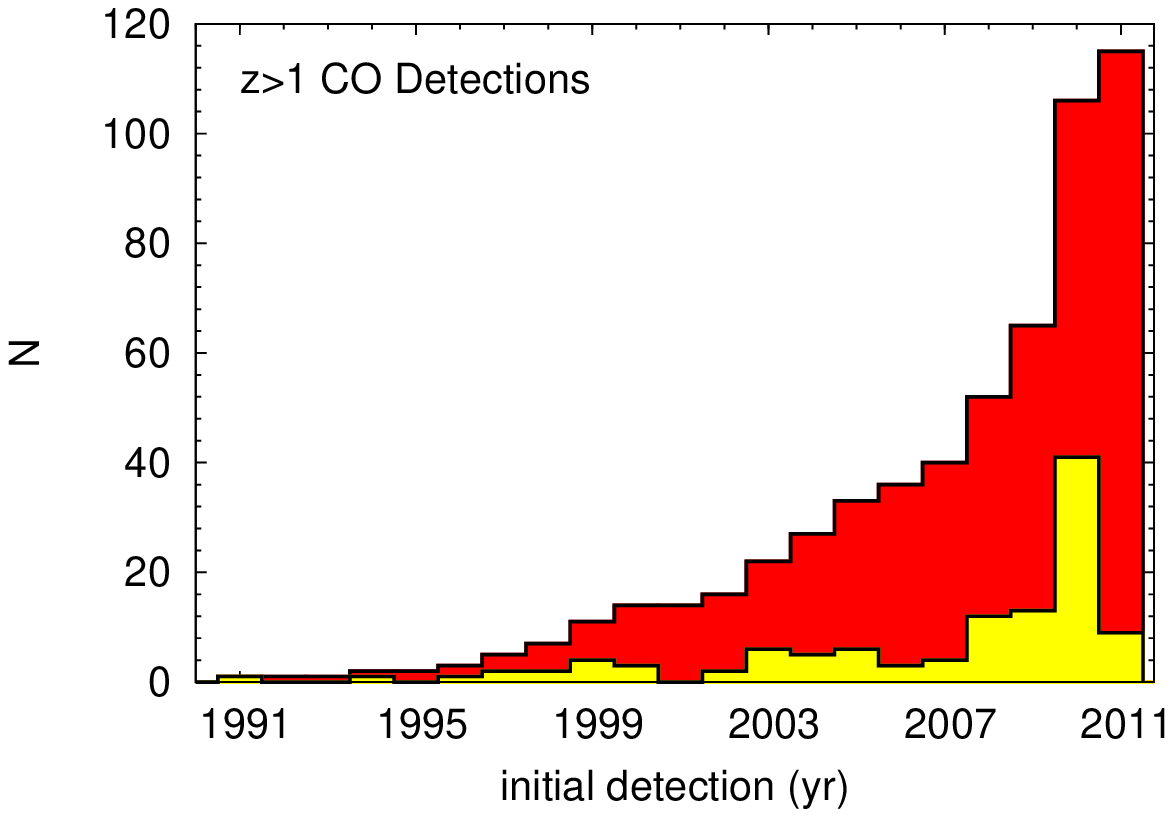}
  \caption{The number of new CO detected galaxies at $z > 1$ per years (yellow). 
Red is the cummulative curve.
}
\end{figure}

More important is the change in the type of galaxies that are being
discovered in CO emission in that last two years. Prior to 2009, the
only galaxies that were detected in CO emission at high $z$ were
extreme starburst galaxies selected in wide field submm surveys (the
submm galaxies, or SMGs), as well as the host galaxies of some very
luminous quasars and radio galaxies. The only exceptions were a few
highly gravitationally magnified Lyman Break Galaxies (LBG; Baker et
al.  2004; Riechers et al. 2010b; Coppin et al. 2007). However, in the
last two years a new class of galaxy has been detected in CO emission:
more typical star forming disk galaxies at $z \sim 1$ to 3 selected
via standard color-color techniques in optical and near-IR deep
fields. These are a hundred times more numerous than the SMGs, and yet
are often as luminous in CO emission.

\section{Extreme starbursts: massive galaxy formation in the
early Universe}

\subsection{General properties}

Molecular gas detections of high redshift galaxies have focused
primarily on extreme starburst galaxies at $z > 1$. These include
highly dust obscured galaxies identified in wide field submm surveys
(SMGs), as well as the host galaxies of optically selected luminous
quasars, and some radio galaxies (Miley \& de Breuck 2008). Typical
submm surveys at 350 GHz detect galaxies at the few mJy level,
implying $L_{FIR} \ge 10^{13}$ L$_\odot$ (ie. 'Hyperluminous infrared
galaxies', HyLIRGs), dust masses $\sim 10^9$ M$_\odot$, and star
formation rates $\ge 10^3$ M$_\odot$ year$^{-1}$. In parallel, about
1/3 of optically selected quasars from eg. SDSS, are detected with
similar FIR luminosities. Solomon \& Vanden Bout (2005) and Blain et
al. (2002) present extensive reviews of these 'Extragalactic Molecular
Galaxies,' and we update some of the information herein.

The areal density of SMGs at $S_{250} \ge 3$mJy is about 0.05 sources
arcmin$^{-2}$ (Bertoldi et al. 2010; Blain et al. 2002). The redshift
distribution has been determined for about 50\% of the SMGs selected
using 1.4GHz continuum observations to determine arcsecond positions.
The median redshift is $z \sim 2.3$, with most of the radio detected
sub-sample within $z \sim 1$ to 3 (Chapman et al. 2003). However,
recently it has become clear that there is a tail of SMGs extending to
high redshift, with possibly 20\% of the sources extending to $z \sim
5$ (Riechers et al. 2010a; Daddi et al. 2009a; 2009b; Schinnerer et al. 
2008; Coppin et al. 2009).

The mean space density of the SMGs is $\sim 10^{-5}$ Mpc$^{-3}$ at
$\sim 2$. (comoving; Blain et al. 2002; Chapman et al. 2003). This
space density is about 1000 times larger than for galaxies of similar
FIR luminosity at $z = 0$, demonstrating the dramatic evolution in the
number density of FIR luminous galaxies with redshift. Daddi et
al. (2009a) conclude, based on SMG space densities and duty cycles,
that there are likely enough SMGs at $z > 3.5$ to account for the
known populations of old massive galaxies at $z \sim 2$ to 3. Study of
the clustering properties of SMGs suggests a minimum halo mass of $3
\times 10^{11}$ M$_\odot$ (Amblard et al. 2011).  However, this
calculation is somewhat problematic due to the low space density, the
broad redshift selection function, and the likely low duty cycle of
SMGs (Chapman et al. 2009).

Michalowski et al.  (2010) present a detailed study of of the radio
through UV SED of SMGs. They find a median stellar mass of $3.7\times
10^{11}$ M$_\odot$. The sources follow the radio-FIR correlation for
star forming galaxies, except perhaps at the highest luminosities,
where a low luminosity radio AGN may contribute. The dust temperatures
span a broad range (10 K to 100 K), with a typical value $\sim 40$K.
They calculate that SMGs contribute about 20\% of the cosmic star
formation rate density at $z \sim 2$ to 4.

\subsection{Molecular gas}

Extensive observations have been performed of the molecular gas in
SMGs and high redshift AGN host galaxies. Typical gas masses derived
from observations of low order CO transitions range from 10$^{10}$
M$_\odot$ to 10$^{11} \times ({\alpha/0.8})$ M$_\odot$ (Hainline et
al. 2006;; Ivison et al. 2011; Riechers et al. 2010a, 2011c; Carilli et
al. 2010; Wang et al. 2010).

\begin{figure}
  \includegraphics[height=70mm]{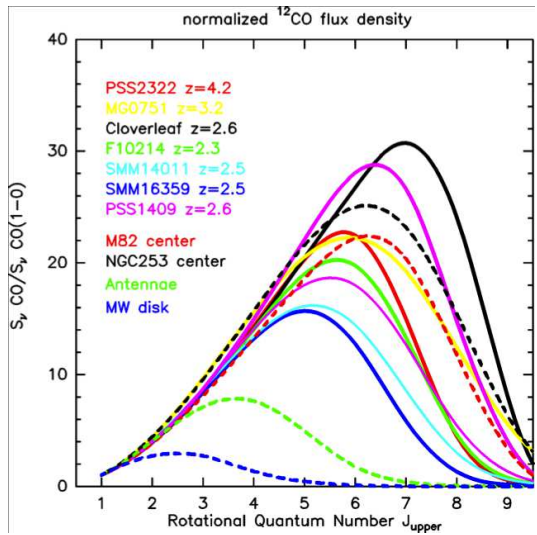}
  \caption{The CO excitation ladder for the integrated emission from
high $z$ SMGs and AGN hosts, plus the dashed lines show the 
inner disk of the Milky Way and selected nearby galaxies 
(from Weiss et al. 2007). The X-axis is the 
upper rotational level of the CO
transition, while the Y-axis is the normalized flux from the CO line.
}
\end{figure}

Figure 6 shows the spectral energy distribution of the CO emission
lines from a number of sources (Weiss et al. 2007). The excitation is
uniformly high, significantly higher than is seen for the CO in the
inner disk of the Milky Way.  The excitation is comparable to what is
found in the nuclear starburst regions on 100pc scales in M82 and NGC
253.  Radiative transfer model fitting to the mean excitation
indicates warm ($\ge 50$ K), dense ($\ge 10^4$ cm$^{-3}$) molecular
gas dominates the integrated CO emission from these galaxies.  Such
conditions are only found in the star forming cores of Giant Molecular
Clouds in the Milky Way on parsec scales. The SMGs show systematically
lower excitation than the QSO host galaxies, and there is mounting
evidence for a lower excitation, more spatially extended molecular gas
distribution in SMGs (Ivison et al. 2011, Harris et al. 2010; Riechers
et al. 2010a; 2011c, Carilli et al. 2010, Papadopoulos et al. 2002;
Scott et al. 2011).

High resolution CO imaging has been performed on a number of SMGs and AGN
host galaxies. Figure 7 shows an example of the CO intensity and
velocity field for the $z = 4.4$ quasar host galaxy, BRI1335-0417
(Riechers et al. 2008). In most cases, the CO emission appears to be
complex, with two or more compact knots of emission separated by a few
kpc (Tacconi et al. 2008; Carilli et al. 2002; Riechers 2008;
2011c). The velocity fields often appear chaotic, with little
indication of rotation, however, this may not be universal for the low
order emission (Carilli et al. 2011).

\begin{figure}
  \includegraphics[height=65mm]{1335.eps}
  \includegraphics[height=55mm]{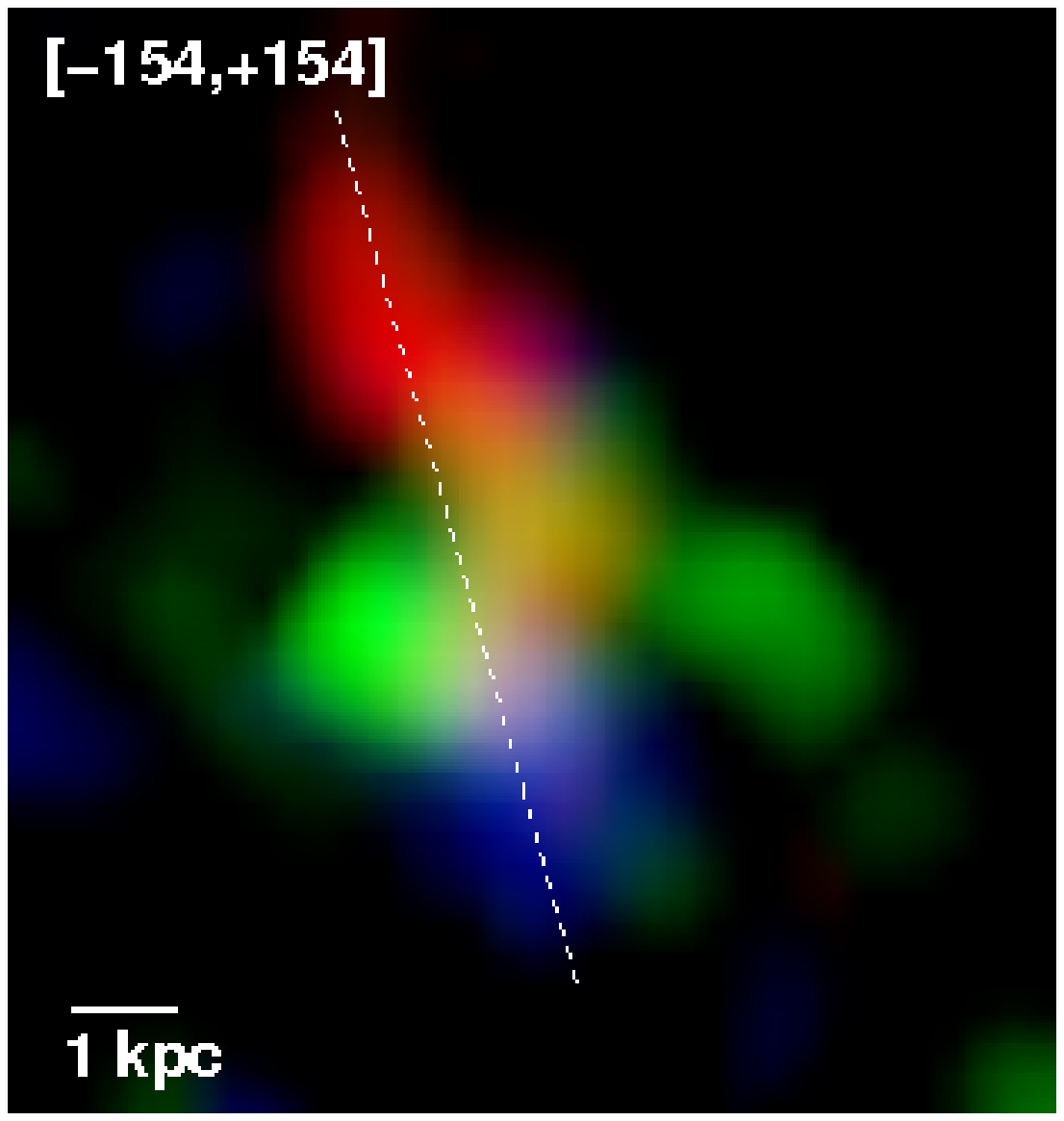}
  \caption{The CO emission from the $z = 4.4$ quasar host galaxy BRI1335-0417
(from Riechers et al. 2008). Left  is the velocity integrated intensity and 
right the mean CO velocity. 
}
\end{figure}

Figure 2 shows the integrated star formation law for both low
redshift, lower luminosity star forming galaxies, and low and high
redshift HyLIRG. There is a non-linear correlation between these two
quantities over a broad range in luminosity.  The relation is
consistent with a powerlaw of index 1.5 between $L'_{CO}$ and
$L_{FIR}$.  These two quantities are linearly related to total
molecular gas mass and star formation rate (Section 2.1). Extensive
analysis has gone into understanding this relationship physically,
both for the integrated correlation, and the spatially resolved
correlation in galaxies (Leroy et al. 2008; Bigiel et al. 2011;
Kennicutt 1998; Krumholz et al. 2009; Narayanan et al. 2011).
Regardless of the physical interpretation and $\alpha$, the empirical
implication of Figure 2 is that the FIR luminosity increases
super-linearly relative to the CO luminosity of galaxies. A review of
the physics of the star formation law is well beyond the scope of this
review. Herein, we make a few simple, empirical points (see also
Section 5.4).

First is that even the AGN sources follow this relationship, which is 
circumstantial evidence for star formation in the host galaxies as the
dominant heating source for the warm dust (see section 4.3). 
And second is that the non-linearity of the relationship implies
shorter gas consumption timescales ($\equiv$ M$_{gas}$/SFR) with
increasing luminosities. For spiral galaxies like the Milky Way, with
$L_{FIR} \sim 10^{10}$ L$_\odot$, the gas consumption timescale is a
few $\times 10^8$ years, while for HyLIRGs this decreases to $\le
10^7$ years, although this depends strongly on the assumed value of
$\alpha$ (see section 5.3).

Progress has also been made on detecting the dense gas tracers, such
as HCO+ and HCN, from high redshift galaxies. These molecules are much
less abundant than CO, but have much higher dipole moments, and hence
stronger rotational transitions. This implies that the radiative
lifetimes are much shorter, and hence maintenance of a Boltzmann
distribution via collisions requires high densities, $\rm n(H_2)
> 10^4$ cm$^{-3}$, even for the lower states, and considerably higher
for the higher order states. Hence, the high order states can be
significantly sub-thermally populated, and emission from these
molecules only comes from the densest molecular gas in
galaxies. Interestingly, $L_{FIR}$ and $L'_{HCN}$ form a linear
correlation (as opposed to the non-linear correlation with $L'_{CO}$; Gao
\& Solomon 2004). The simplest interpretation is that observations of
these high density tracers simply 'count' star forming clouds in
galaxies, and that the properties of the dense clouds are relatively
universal (Wu et al. 2005).

The emission lines from the dense gas tracers are typically an order
of magnitude weaker than CO, although the ratio varies dramatically
between galaxies (Gao \& Solomon 2004). A few galaxies have been
detected in the dense gas tracers at high redshift, predominantly
strongly gravitationally lensed systems (Riechers et al. 2011a,b; Wagg
et al.  2005, Carilli et al. 2005; Solomon et al. 
2003).  The high redshift galaxies
generally follow the low redshift correlations. Interestingly, in some
cases even the high order transitions are excited, indicating either
extremely dense gas, or a contribution to the excitation by the AGN
(Riechers et al. 2011a,b; Wagg et al. 2005).

\subsection{Topics on quasar hosts}

\subsubsection{Dust and molecular gas in the most distant galaxies}

Detection of molecular line emission at the very highest redshifts ($z
> 6$) has thus far been limited to quasar host galaxies.  These
galaxies generally follow the trends discussed above for SMGs and
lower redshift quasar hosts in terms of their warm dust and molecular
gas properties, and in particular, the 1/3 fraction of submm
detections of quasar host galaxies at $S_{250GHz} \ge 2$mJy remains
constant to the highest redshifts, implying HyLIRG host galaxies.

\begin{figure}
  \includegraphics[height=65mm]{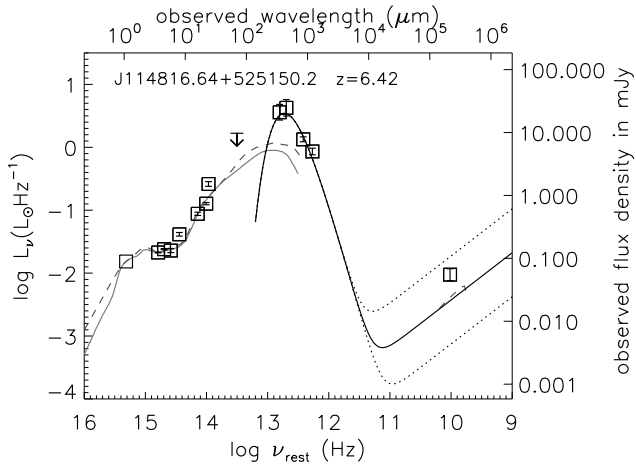}
  \caption{The UV through radio SED of the $z = 6.42$ quasar J1148+5258
(from Wang et al. 2008). The curves in the UV through mid-IR are local quasar
templates from Elvis et al. (1994) and  Richards et al. (2006). The
curves in the Far-IR through radio are for a 50K dust model that
obeys the radio-FIR correlation for star forming galaxies. 
}
\end{figure}

Figure 8 shows the SED of the most distant SDSS quasar J1148+5258 from
UV to radio wavelengths. This SED is typical of the submm detected
high $z$ quasars (Wang et al. 2009, 2010; Leipski et al.
2010).  From the UV through mid-IR
($\sim 30\mu$m rest frame), the SED is consistent with that of 
lower redshift SDSS quasars, including a hot dust (1000K) seen in
the mid-IR, heated by the AGN.
However, the submm detected sources show a
substantial excess over the lower $z$ SDSS quasar SED. This
excess is well fit by a warm dust component ($\sim
50$K). Extrapolating this component to the radio also shows that most
of the sources follow the radio-FIR correlation for star forming
galaxies. These results argue that
the warm dust component is heated by star formation in the host
galaxy. The star formation rates are $\sim 10^3$ M$_\odot$
year$^{-1}$, implying a major starburst coeval with the AGN in the
host galaxy.

CO has been detected in every $z \sim 6$ quasar host galaxy that was
selected via a previous submm detection of the dust.  To date, 11
quasar host galaxies have been detected in CO emission between $z =
5.7$ and 6.4 (Wang et al. 2010), with gas masses $\ge 10^{10}$
M$_\odot$.

The detection of large dust masses within 1Gyr of the Big Bang
immediately raises an interesting question: how does so much dust form
so early?  One standard dust formation mechanism in the ISM involves
coagulation in the cool winds from low mass stars, which, naively
would take too long. The large dust masses have led to a number of
theoretical studies of early dust formation, with models involving:
dust formation associated with massive star formation in eg.
supernova remnants (Dwek et al. 2007; Venkatesen et al 2006), dust
formation in outflows from the broad line regions of quasars (Elitzur
in prep; Elvis et al. 2002), and dust formation in the gas phase ISM
(Draine 2003). Michalowski et al. (2010) consider this problem in
detail, and show that AGB stars are insufficient, and even SNe require
a very top-heavy IMF and unrealistic dust yields.

Recent observations of the UV-extinction curves in a few $z \sim 6$
quasars and GRBs suggest a different dust composition at $z > 6$ 
relative to the Milky Way or the SMC, as well as
relative to quasars at $z < 4$.  The extinction can be modeled by
larger silicate and amorphous carbon grains (vs. eg. graphite), as
might be expected from dust formed in supernova remnants (Stratta et
al. 2007; Perley et al. 2010). The formation of dust in the early
Universe remains an interesting open question.

\subsubsection{Fine structure lines: [CII] 158$\mu$m}

At high redshift the FIR atomic fine structure lines are observed in the
submm band, and hence can be studied with existing ground-based
telescopes. In particular, substantial progress has been in the study
of [CII] in distant galaxies.

We have started a systematic search for [CII] emission from $z > 4$
quasars quasar host galaxies (Maiolino et al. 2005;
Wagg et al. 2010). At the highest redshifts, we now have three
detections of the [CII] line from $z > 6.2$ quasar host galaxies
(Bertoldi et al. in prep).

Figure 9 shows the [CII]  images of the $z = 6.42$ quasar
J1148+5258, as well as the mm continuum and VLA CO 3-2 images, at
0.25$"$ resolution from the PdBI (Walter et al. 2009). The [CII]
emission is extended over about 1.5 kpc, while the CO is even more
extended. If [CII] traces star formation, the
implied star formation rate per unit area $\sim 10^3$ M$_\odot$
year$^{-1}$ kpc$^{-2}$. This value corresponds to the predicted upper
limit for a 'maximal starburst disk' by Thompson et al. (2005), ie. a
self-gravitating gas disk that is supported by radiation pressure on
dust grains. Such a high star formation rate areal density has been
seen on pc-scales in Galactic GMCs, as well as on 100 pc scales in the
nuclei of nearby ULIRGs. For J1148+5251 the scale for the disk is yet
another order of magnitude larger.

\begin{figure}
  \includegraphics[height=60mm]{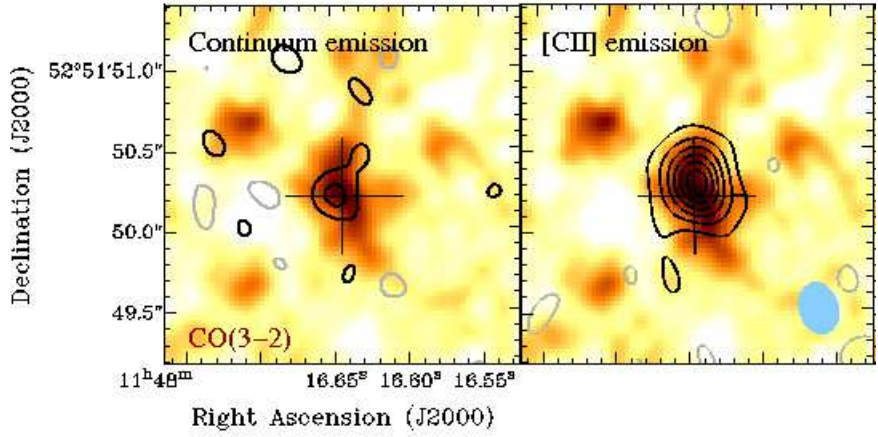}
  \caption{The images of the $z = 6.42$ quasar
J1148+5251. Images from the PdBI show the dust and [CII] emission (contours, left and right,
respectively) at $0.25"$ resolution, plus the VLA CO 3-2 in color 
(from Walter et al. 2009; 2004). }
\end{figure}

One potential difficulty with using the [CII] line as a star formation
diagnostic is the very broad range in the ratio of [CII] to Far-IR
luminosity, in particular for galaxies with $L_{FIR} > 10^{11}$
L$_\odot$. Stacey et al. (2011) present an analysis of this ratio
versus $L_{FIR}$ for low and high redshift galaxies.  At high
luminosity, the distribution is essentially a scatter plot, with the
ratio ranging by 3 orders of magnitude, although there appears to be
less scatter if AGN are removed. Figure 10 shows this ratio versus
dust temperature. Malhotra et al.  (2001) consider temperature to be
the more fundamental quantity, due to inefficiency of photoelectric
heating of charged dust grains in high radiation environments. Figure
10 shows a possible correlation of the [CII] to $L_{FIR}$ ratio with
dust temperature, but again, the scatter is very large. Papadopoulos
et al. (2010) also point out that dust opaticy might play a role in
decrease the [CII] luminousity from extreme starburst galaxies.

\begin{figure}
  \includegraphics[height=75mm]{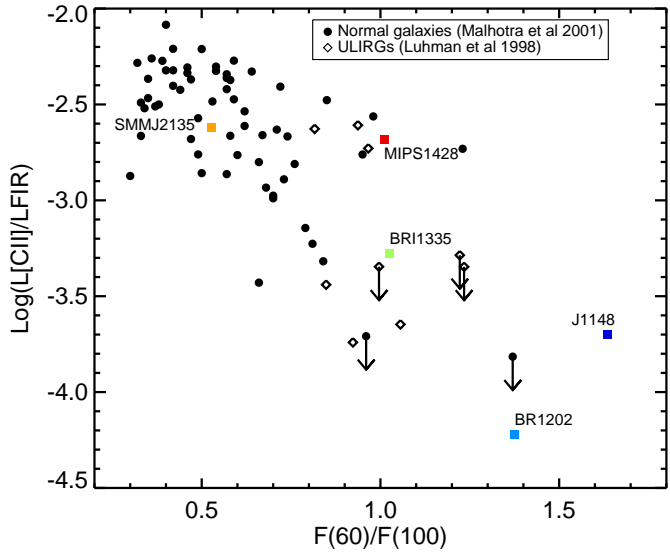}
  \caption{The [CII]/$L_{FIR}$ ratio versus
dust temperature for low and high redshift galaxies (Wagg et al. in prep.). }
\end{figure}

\subsubsection{The Black Hole -- Bulge mass relation within 1 Gyr of
the Big Bang}

There is a well studied correlation between the masses of black holes
at the centers of galaxies, and the velocity dispersion of the host
galaxies. This M$_{BH}$ - $\sigma_V$ relation 
implies a roughly linear correlation between
black hole and spheroidal galaxy mass, with a proportionality constant
of: $\rm M_{BH} = 0.002 M_{bulge}$. This correlation has been used to
argue for a 'causal connection between the formation of supermassive
black holes and their host spheroidal galaxies' (Gebhardt et al. 2000;
Kormendy \& Bender 2011; H\"aring \& Rix 2004; Gultekin et al. 2009). 
While AGN feedback via
winds or jets has been invoked to explain the effect, the details
remain obscure.

\begin{figure}
  \includegraphics[height=70mm]{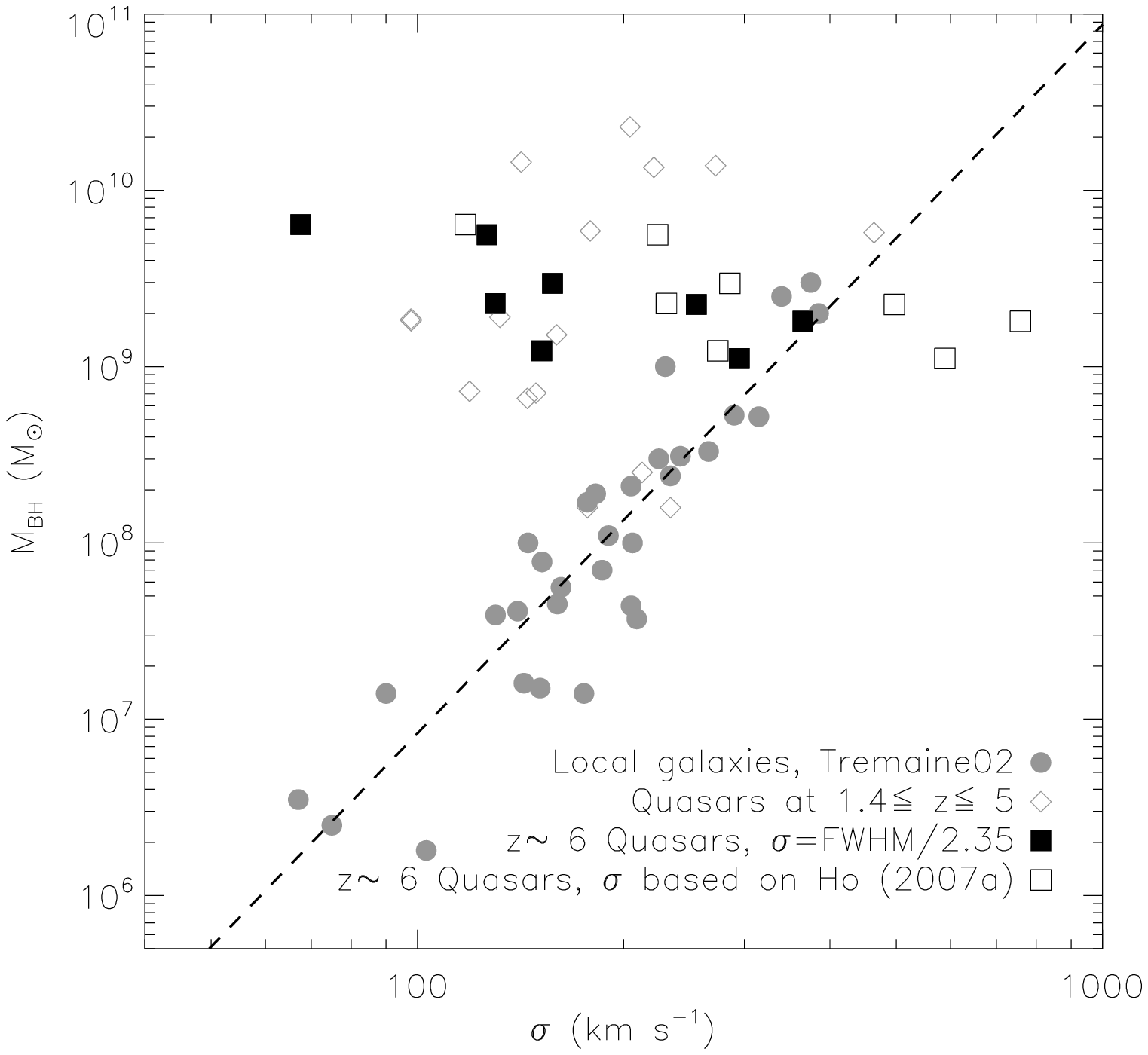}
  \caption{The relationship between the black hole mass and the host 
spheroidal galaxy velocity dispersion for low and high redshift quasars (from Wang
et al. 2010a). For the high $z$ sources the velocity dispersion is estimated
from the CO line imaging. 
}
\end{figure}

High resolution imaging of gas dynamics allows for study of the
evolution of the black hole -- bulge mass relation to high redshift
redshift.  Imaging of a few $z \ge 4$ quasars shows a systematic
departure from the low $z$ relationship (eg. Walter et al. 2004;
Riechers et al. 2008). Figure 11 shows a compilation by Wang et
al. (2010). They find that, assuming random inclination angles for the
molecular gas, the $z \sim 6$ quasars are, on average, a factor 15
away from the black hole -- bulge mass relation, in the sense of
over-massive black holes.  Alternatively, all of the $z \sim 6$
quasars could be close to face-on, with inclination angles relative to
the sky plane all $< 20^o$.  High resolution imaging of the CO
emission from these systems is required to address the interesting
possibility that the black holes form before the host spheroids.

\subsection{Massive galaxy formation at high redshift}

Overall, the observations of the molecular gas and dust in extreme
starburst galaxies at high redshift (SMGs, AGN hosts)
indicate a major star formation episode during the formation
of massive galaxies in group or cluster environments.

A key question is: what drives the prolific star formation?  Tacconi
et al. (2006; 2008) argue, based on imaging of higher-order CO
emission from a sample of $z \sim 2$ SMGs, that SMGs are predominantly
nuclear starbursts, with median sizes $< 0.5"$ ($< 4$kpc),
`representing extreme, short-lived, maximum star forming events in
highly dissipative mergers of gas rich galaxies.'  This conclusion is
supported by VLBI imaging of the 1.4 GHz emission from star forming
regions in two SMGs (Momjian et al. 2005; 2010). 

However, recent EVLA imaging of the lower order CO emission in 
SMGs (Ivison et al. 2011; Carilli et al. 2010; Riechers et al. 2011c),
suggests that the lower-excitation molecular gas reservoirs can be
significantly more extended. Riechers et al. (2011c) suggest a
sequence in which the SMG phase is an early stage of a major gas rich
merger, with the quasar phase arising later in the evolution (Sanders 
et al. 1988).

\section{Secular galaxy formation during the epoch of galaxy assembly}

\subsection{sBzK and other typical star forming galaxies at $z \sim
1$ to 3}

Optical through near-IR color selection techniques have become
remarkably efficient at finding both star forming and passive galaxies
during the peak epoch of galaxy assembly ($z \sim 1$ to 3). These
include (rest frame) UV-selected samples (BX/BM; Steidel et al. 2004)
and near-IR selected samples (sBzK; Daddi et al. 2005). Grazian et
al. (2003) analyze the substantial overlap between the populations.
The critical aspect for these samples is that they are not rare,
pathological galaxies, such as luminous quasar hosts or the
hyper-starburst submm galaxies. These galaxy samples are generally
representative of the broad distribution of star forming galaxies at
these epochs (Section 1), with areal densities of a few
arcmin$^{-2}$, or volume densities $\ge 10^4$ Mpc$^{-3}$.

\begin{figure}
  \includegraphics[height=57mm]{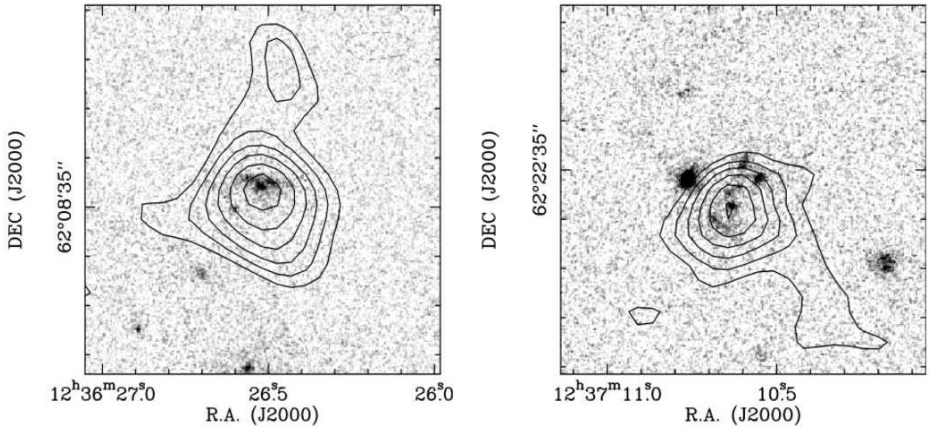}
  \caption{The CO emission from two $z \sim 1.5$ sBzK galaxies,
with two CO velocity intervals shown as contours, and the greyscale is the
HST i-band image (from Daddi et al 2010a).
}
\end{figure}

HST imaging of these galaxies show clumpy but predominantly disk-like
galaxies with sizes of order 10 kpc (Figure 12; Daddi et al. 2010a),
with stellar masses $\ge 10^{10}$ M$_\odot$.  Imaging of the H$\alpha$
and CO emission, reveals turbulent, but systematically
rotating gas disks, and giant star forming clumps up to 1kpc in size
(Genzel et al. 2008, 2011; F\"orster-Schreiber et al. 2011; Tacconi et
al. 2010).

\subsection{COSMOS radio stacking: dawn of downsizing}

Pannella et al. (2010) have selected a sample of 30,000 sBzK galaxies
from the COSMOS survey to determine the mean dust-unbiased star
formation rates using stacking of 1.4 GHz observations.  The number of
sources, and the sensitivity of the radio observations, allow for
substantial binning of the galaxies as a function of stellar mass,
star formation rate, color, and blue magnitude.  There is no
correlation between median star formation rate and blue
magnitude. This lack of correlation of SFR with blue magnitude occurs
because there is a strong correlation of star formation rate with B-z
color, ie. the extinction increases with star formation rate. There is
also a positive correlation of star formation rate with stellar mass.

Combining these data, Figure 13 shows the specific star formation rate
(SFR/M$_{stars}$) versus stellar mass for the COSMOS sBzK sample.  For
comparison, the relation at $z = 0.3$ from Zheng et al. (2007) is also
shown. A line shows the inverse Hubble time at $z = 1.8$.
Galaxies above this line have SSFR that are sufficient to form the
observed stars in the galaxy over their Hubble time. Galaxies below
this line required a substantial increase in SFR rate in the past to
form the stars that are seen. Lastly, the open points showing the SSFR
based on dust-uncorrect star formation rates from the UV measurements.

\begin{figure}
  \includegraphics[height=80mm]{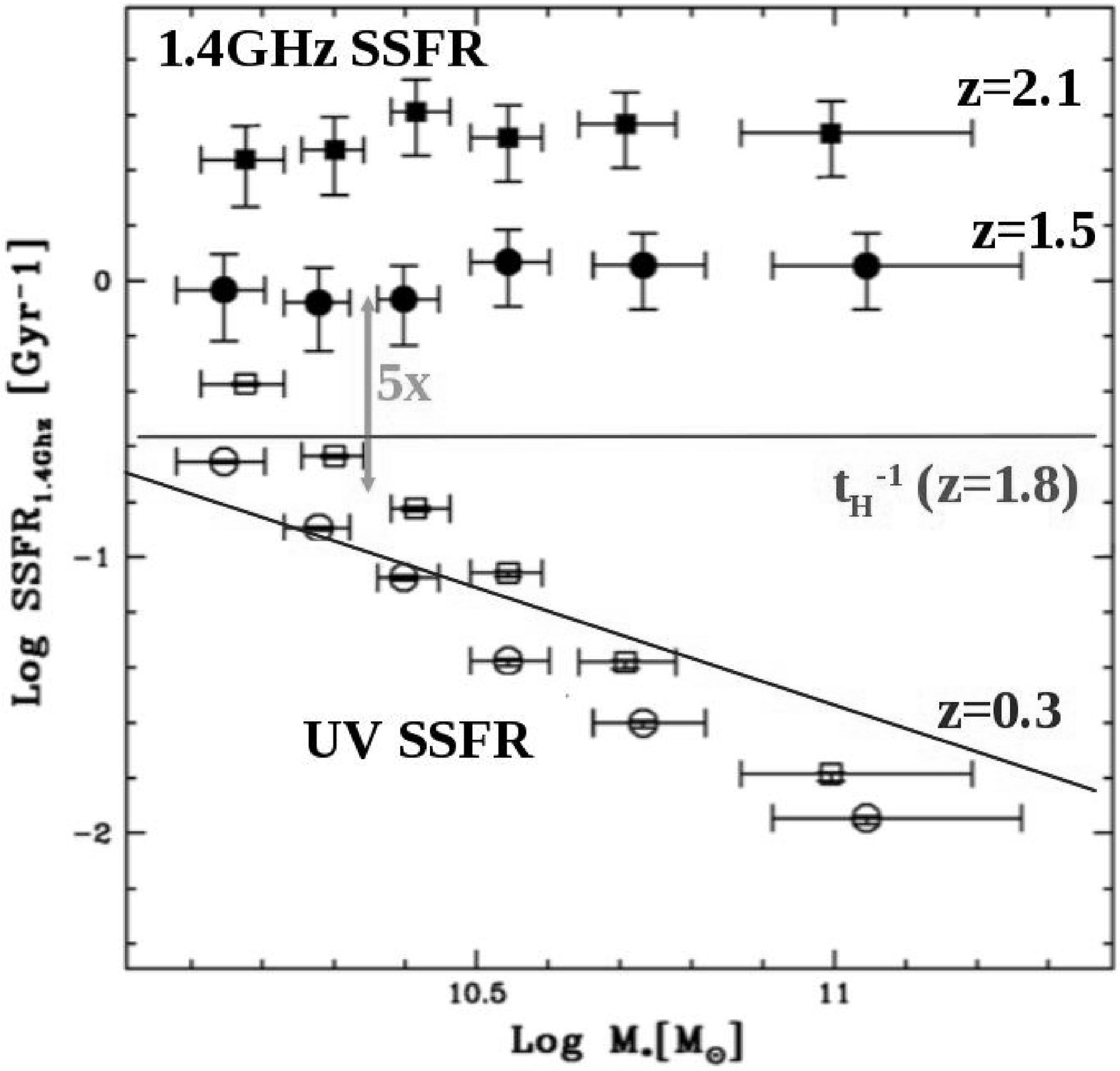}
  \caption{The specific star formation rate for $z \sim 2$ 
COSMOS sBzK galaxies (Pannella et al. 2009).  Black filled symbols are
SFR results from stacking of 1.4 GHz emission. Open symbols are the 
UV derived SFR without an extinction correction. A line
indicates the SSFR vs. stellar mass for $z \sim 0.3$ galaxies in the
Zheng et al. (2007) sample.
}
\end{figure}

From Figure 13, the SSFR increases with redshift, even between $z =
1.5$ and 2.1.  The high redshift galaxies are all above the 'red and
dead' line indicated by the inverse Hubble time.  Pannella et al. find
that the substantial negative slope with stellar mass seen at $z =
0.3$ (ie. the decrease in SSFR with increasing stellar mass) becomes
essentially flat at $z \ge 1.5$. Hence, $z \sim 2$ corresponds to an
epoch when even fairly massive but common galaxies are actively
forming stars. Lastly, it is clear that the dust extinction increases with
increasing star formation rate. Interestingly, the standard
factor 5 UV extinction correction for LBGs occurs at a stellar mass of
$\sim 2\times 10^{10}$ M$_\odot$, which is typical of LBG samples
(Shapley et al. 2003).

\subsection{Molecular gas: gas-dominated galaxies}

Perhaps the most interesting result from the study of the sBzK and
BX/BM galaxy samples comes from the searches for molecular gas.  These
samples show a remarkably high detection rate ($> 50\%$), in CO emission
(Daddi et al. 2010a; Tacconi et al. 2010). The line strengths are
comparable to those seen in SMGs and quasar hosts, but the star
formation rates are an order of magnitude smaller. Figure 12 shows
some examples.  High resolution imaging shows that the CO is extended
on the same scale as the optical disks, ($\sim 10$kpc), with large
condensations of size $\sim 1$kpc, and masses $> 10^9$ M$_\odot$
(Tacconi et al.  2010; Aravena et al. 2010; Daddi etal. 2009a).

Some of these galaxies have been observed in the CO 1-0 transition
with the VLA, as well as higher order transitions with the PdBI
(Figure 14; Dannerbauer et al. 2009; Aravena et al. 2010).  The
excitation up to CO 3-2 appears to be sub-thermal, and substantially
lower than is seen in either quasar hosts and SMGs (Figure 6).  The
excitation up to CO 3-2 is comparable to the Milky Way disk.
Likewise, Figure 15 shows that the ratio of CO luminosity to FIR
luminosity in these galaxies is similar to the Milky Way, and not to
the SMG population, or compact nuclear starbursts.

\begin{figure}
  \includegraphics[height=70mm]{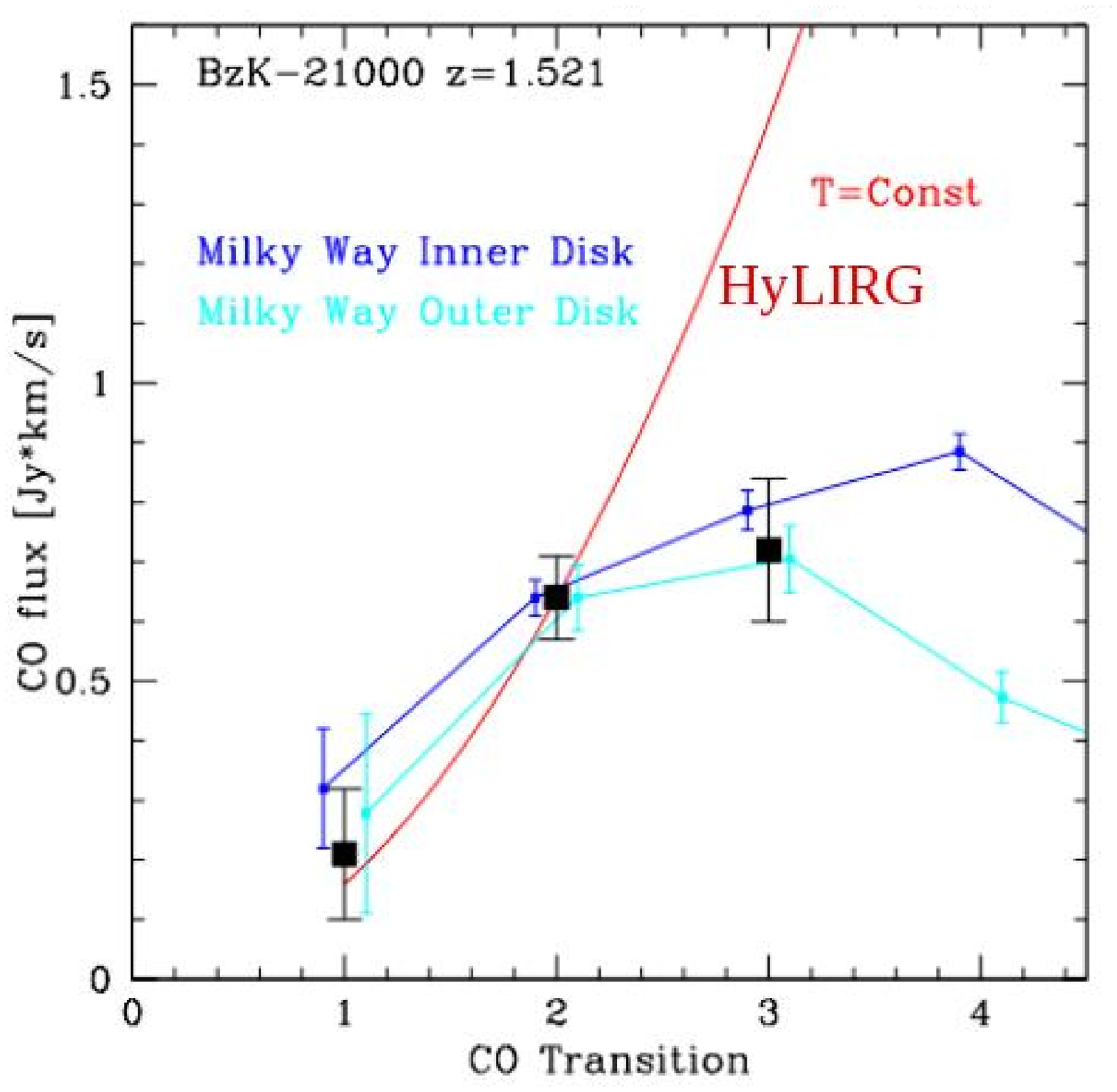}
  \caption{The CO ladder for sBzK galaxies (CO flux versus rotational
quantum number for the upper state. The redline indicates constant brightness
temperature (flux $\propto \nu^2$), typical for SMGs and AGN hosts in
this J range (Dannerbauer et al. 2009; Aravena et al. 2010). 
}
\end{figure}

Daddi et al. (2010a) have done a detailed analysis of the possible CO
luminosity to H$_2$ mass conversion for the sBzK sample. They employ
dynamical models of forming disk galaxies including dark matter, as
well as extensive observations of the stellar content and the CO
dynamics. They conclude that the CO conversion factor is likely
similar to the Milky Way value, rather than the nuclear starburst
value employed for SMGs and quasar hosts. This conclusion is
consistent with the Milky Way-like excitation and $L'{CO}/L_{FIR}$
ratio, and the large scale for the CO disks. Tacconi et al. (2010)
reach a similar conclusion on $\alpha$ for the BX/BM samples.

\begin{figure}
  \includegraphics[height=70mm]{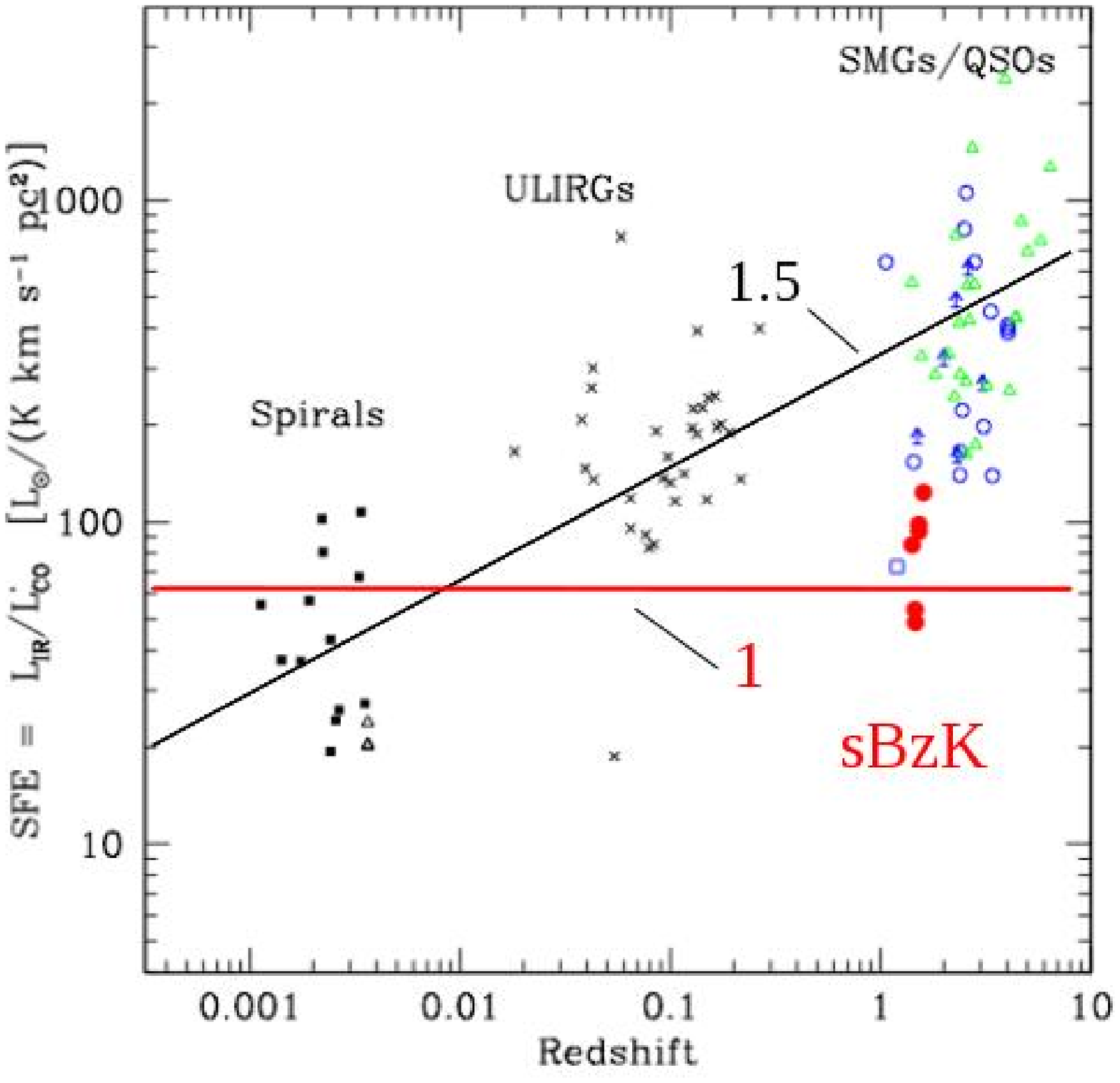}
  \caption{The ratio of FIR to CO luminosity for sBzK galaxies (red), and
other galaxy samples (from Daddi et al. 2010a). The lines indicate powerlaw 
relationships of  different indices. 
}
\end{figure}

The implied H$_2$ masses are then of order 10$^{11}$ M$_\odot$. The
gas masses are comparable to, or larger than stellar masses in 
the sBzK and BX/BM samples (Daddi et al. 2010a; Tacconi et al. 2010). 
This is very different with respect to low redshift disk
galaxies, where the baryon content is dominated by stars (Figure
16). Hence, the peak epoch of cosmic star formation also
corresponds to an an epoch when the dominant baryon component 
in star forming galaxies is gas, not stars.

\begin{figure}
  \includegraphics[height=80mm]{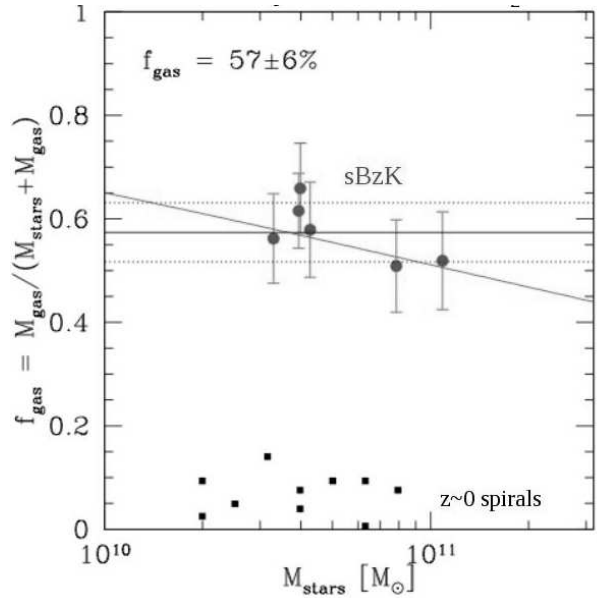}
  \caption{The molecular gas baryon fraction for sBzK galaxies at $z \sim 1.5$
(circles), and low redshift spirals (squares; from Daddi et al. 2010a)).
}
\end{figure}

\subsection{Normal galaxy formation in the gas rich-era}

The implied gas consumption timescales for the sBzK and BX/BM galaxies
is $\sim \rm few \times 10^8$ years.  This is an order of magnitude
longer than the gas consumption timescale for the hyperstarbursts in
SMGs and quasar host galaxies. Genzel et al. (2010) consider this
point in detail, and conclude that the most likely explanation relates
to different global dynamical effects in disks versus compact
starbusts.  They also emphasize that the gas consumption timescales
are much shorter than the Hubble time in either case, and hence star
formation must be a balance between gas accretion and feedback (see
also Bauermeister et al. 2010).

Theoretical and numerical studies have recently suggested that the
dominant mode of star formation in the Universe, again peaking around
$z \sim 1$ to 3, is not related to major gas rich merger events, but
is driven by the slower process of 'cold mode accretion' (Dekel et
al. 2009; Keres et al. 2009).  In the CMA model, gas flows into
galaxies from the IGM along cool, dense filaments. The flow never
shock-heats due to the rapid cooling time, but continuously streams
onto the galaxy at close to the free-fall time. This gas forms a
thick, turbulent, rotating disk which efficiently forms stars across
the disk, punctuated by giant clouds of enhanced star formation on
scales $\sim$ few kpc. Genzel et al. (2008; 2011) show that the
process is consistent with marginally  Toomre-unstable gaseous disks.
These star forming regions then migrate to the galaxy center via
dynamical friction and viscosity, forming compact stellar bulges
(Genzel et al. 2008; Bournaud et al. 2009; Elmegreen et al. 2009).
The process is regulated by feedback, both within the giant star
forming clouds themselves, and possibly from an active nucleus (Dave
et al. 2011; Genzel et al. 2010).

The CMA process leads to relatively steady and active ($\sim 100$
M$_\odot$ yr$^{-1}$) star formation in galaxies over timescales
approaching 1 Gyr.  The process slows down dramatically as gas supply
decreases, and the halo mass increases, generating a virial shock in
the accreting gas.  Subsequent dry mergers at lower redshift then lead
to continued total mass build up, and morphological evolution, but
little subsequent star formation.  

The H$\alpha$ and CO dynamical analyses of these samples are generally
consistent with the morphologies expected from CMA, including
turbulent but systematically rotating, gas disks, and giant star
forming clumps.  Shapiro et al. (2008) show that such a study of disk
kinemetry enables an 'empirical differentiation between merging and
non-merging systems'. Of course, this remains just a consistency
check, and not direct observation of CMA.

Interestingly, observations of even more luminous star forming
galaxies at even high redshift ($z > 4$) suggest that CMA may scale up
to SMG lumnosities at the highest redshifts (Carilli et al. 2010).

\section{Dense gas history of the Universe}

While admittedly compressing much information, the star formation
history of the Universe has been a dominant tool in the study of
galaxy formation over the last decade (Figure 1). However, the
relationship between star formation and the molecular gas
content of galaxies (Figure 2), implies that the SFHU should be
reflected in the evolution of molecular gas. 

\begin{figure}
  \includegraphics[height=70mm]{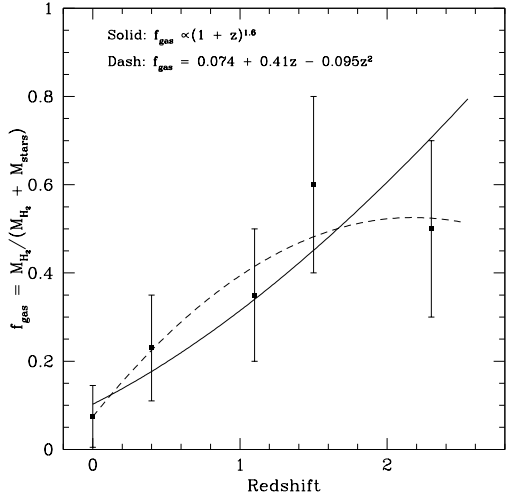}
  \caption{The molecular gas baryon fraction for star forming galaxies
with M$_{stars} \ge 10^{10}$ M$_\odot$ (adapted and revised from Daddi et al. 
2010a; Geach et al. 2011; Bauermeister et al. 2010). 
}
\end{figure}

Section 5.3 discusses how the average molecular gas content of star
forming galaxies rises substantially with redshift. This is quantified
in Figure 17, which shows the mean molecular gas baryon fraction
($\equiv \rm M_{gas}/[M_{gas} + M_{stars}]$), for star forming
galaxies with $\rm M_{stars} > 10^{10}$ M$_\odot$ (adapted from Daddi
et al. 2010b; Geach et al. 2011; Bauermeister et al. 2010). Admittedly
there are many selection effects and assumptions that enter this
calculation, but the current data support the idea of a substantial
increase in the molecular gas content of galaxies at the peak epoch of
cosmic star formation.

\begin{figure}
  \includegraphics[height=80mm]{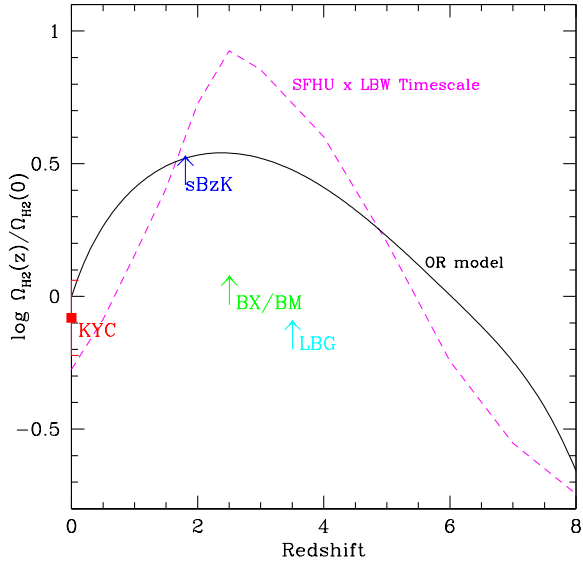}
  \caption{Gross estimates of the cosmic density (comoving) of molecular 
gas versus redshift based on the samples of star forming
galaxies observed at $z > 1$. These include the near-IR selected sBzK
galaxies (Daddi et al. 2010a), optically selected (BX/BM) galaxies
(Tacconi et al. 2010; Genzel et al. 2010), and
LBGs (Riechers et al. 2010b). Also plotted are two basic models: 
 one boot-strapped from
the SFHU plot assuming a simple timescale for gas conversion to stars
in galaxies (Leroy et al. 2008). The second is based on
the work of Obreschkow and Rawlings (2009) and the Millenium simulations. 
The zero redshift point is from the CO survey of Keres et al. (2003).}
\end{figure}

The next level of abstraction is to sum the gas mass to obtain the
evolution of the cosmic density of molecular gas. Figure 18 shows an
initial, very gross attempt at such a calculation, based on the 115 CO
detections at $z > 1$ to date. The gas mass calculation entails a
simple scaling from stellar mass densities from Grazian et al. (2007)
using the mean gas baryon fraction for the different samples (Daddi et
al. 2009b; Tacconi et al. 2010; Riechers et al. 2010b). We include the
$z = 0$ value from Keres et al.  (2003). This figure includes two
basic models: one boot-strapped from the SFHU plot assuming a constant
timescale for gas conversion to stars in galaxies (Bigiel et al. 2011;
Leroy et al. 2008; Bauermeister et al. 2010). The second is based on
the work of Obreschkow and Rawlings (2009), who predict the molecular
gas content of galaxies based on the Millenium simulations.

We emphasize this plot remains highly speculative due to the many
assumptions involved, in particular, the conversion factor of CO
luminosity to molecular gas mass, the open question of the evolution
of the star formation law, and very limited samples. 

\section{ALMA and EVLA}
 
The last decade has seen the opening of the high $z$ Universe to radio
observations of the dust, gas, and star formation
in the first galaxies. These results emphasize the critical need for
pan-chromatic studies of galaxy formation to reveal all the key elements
of the complext process. Fortunately, the immediate promise for a
dramatic improvement in these studies is being realized with the
science commissioning of the Expanded Very Large Array and the
Atacama Large Millimeter Array.  

The EVLA is a complete reinvention of the VLA, building on existing
infrastructure (telescopes, rail track), but completely replacing the
entire (1970's) electronic systems, from receivers through LO/IF to
the correlator, to establish an essentially completely new telescope
for the coming decade (Perley et al. 2011). The EVLA has complete
frequency coverage from 1 GHz to 50GHz, and the bandwidth has
increased by a factor 80, to 8GHz with thousands of spectral channels,
greatly improving capabilities for spectral line searches and
studies. The continuum sensitivity has increased by up to an order of
magnitude. The array still provides compact configurations to image
larger structures, and extended configurations that provide
resolutions down to 40mas at 40GHz. 

The EVLA construction project is close to completion, and the full
array is operating at $\ge 18$ GHz with up to 2 GHz of bandwidth. 
Early science observations have been proceeding since
March 2010, and indeed, some of the results discussed above are based
on these observations.

The ALMA telescope will consist of 54 12m antennas for sensitive
observations, plus 12 7m antennas for wide field imaging (Wootten \&
Thompson 2009), located at one of the best submm observing sites in the
world, at 5000m elevation in Chile. The array will work from 80GHz to
750 GHz, initially in four bands, with an instantaneous bandwidth of
8GHz. The array configurations can be adjusted for wide field imaging
as well as high resolution imaging to 20mas resolution at 350 GHz.
Figure 4 shows the sensitivity of ALMA compared to existing (sub)mm
arrays. The 3 orders of magnitude improvement in sensitivity in the
submm is particularly dramatic.

The ALMA is also well into construction, with about half the antennas
either operating at the high site, or under construction at the
observational support facility. Demostration science observations have
already shown the power of even a limited set of ALMA antennas to
perform ground-breaking observations of the dust and gas in galaxies,
and early science observations will start at the end of 2011.

Taken together, ALMA and the EVLA represent an order of magnitude, or
more, improvement in observational capabilities from 1 GHz up to 1
THz. Such a jump in capabilities is essentially unprecedented in
ground-based astronomy. Following are a few examples of 
the potential of these instruments. 

Figure 19 shows the sensitivity of ALMA, and other telescopes, to the
[CII] line emission versus redshift. Included are the expected signals
of star forming galaxies of different luminosity. ALMA will detect the
[CII] line from typical LBGs well into cosmic reionization.  Indeed, the
8GHz bandwidth and sensitivity present the interesting potential for
determining the redshifts for these first galaxies.  Determining
optical spectroscopic redshifts is particularly hard as the Ly$\alpha$
line shifts into the near-IR, and the presence of a partially neutral
IGM can attenuate the Ly$\alpha$ emission as well.

\begin{figure}
  \includegraphics[height=70mm]{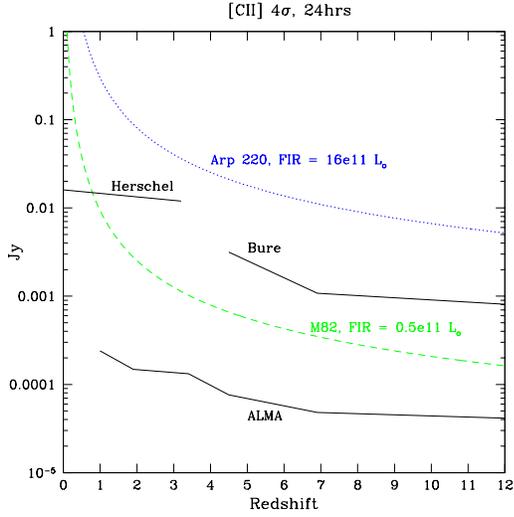}
  \caption{The expected [CII] line peak flux density versus redshift for 
active and dwarf star forming galaxies, plus the sensitivity of ALMA 
and other instruments. Note that there are gaps in the frequency coverage
due the atmostphere and availability of receivers that are not depicted.}
\end{figure}

Figure 20 shows what an 8GHz spectrum of a luminous starburst, like
J1148+5258 at $z = 6.42$, might look like with ALMA. Numerous interesting
transitions of important diagnostic molecules will be detected, including
dense gas tracers, isotopes, and isomers. ALMA opens up full astrochemical
studies of the first galaxies. 

\begin{figure}
  \includegraphics[height=60mm]{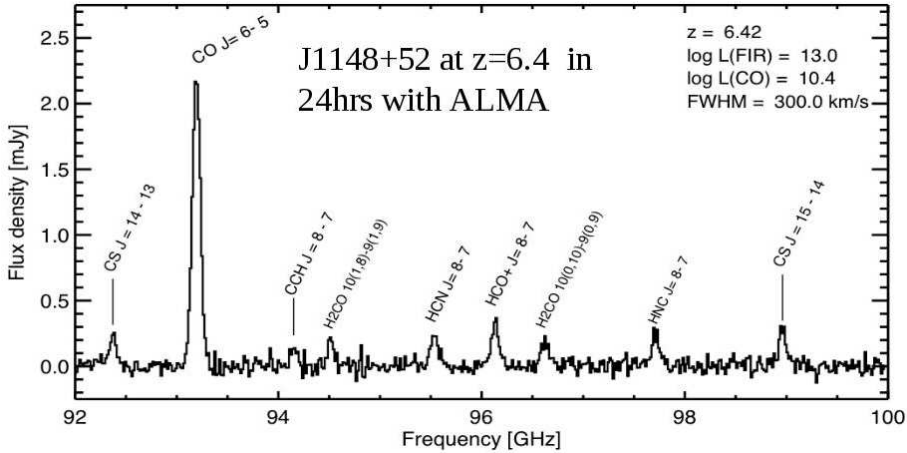}
  \caption{A simulated spectrum of the HyLIRG J1148+5258 at $z = 6.42$ 
with 8GHz using ALMA in 24 hours. 
}
\end{figure}

The EVLA has a 25\% fractional bandwidth at 30 GHz. Hence, every
observation with the EVLA at high frequency will include a
blind search for molecular line emitting galaxies at high
redshift. For instance, an EVLA observation at 19 to 27 GHz covers CO
1-0 at $z = 3.2$ to 5.0 instantaneously, thereby obviating the need
for optical spectroscopic redshifts.  Figure 21 shows a recent example
from EVLA early science of this potential.  Three molecule rich
galaxies have been observed in a single pointing and 256MHz bandpass
with the EVLA.  Essentially every long observation of the EVLA over 20GHz
will detect molecular line emission from distant galaxies,
whether intended or not.

\begin{figure}
  \includegraphics[height=100mm]{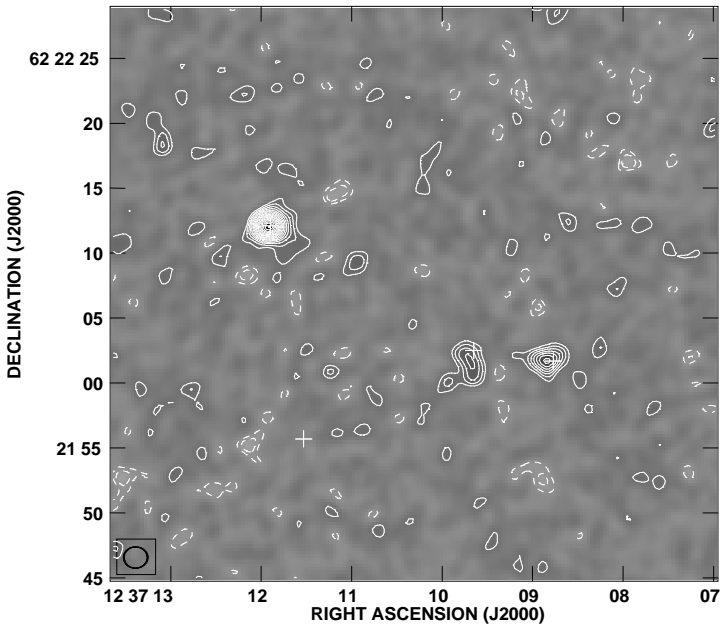}
  \caption{The GN20 field seen with the EVLA with a 256 MHz bandwidth at 
45 GHz. Three galaxies are detected in CO 2-1 emission at $z = 4.0$ (from 
Carilli et al. 2011). 
}
\end{figure}

\acknowledgements CC thanks the Astronomische Gesellschaft organizers
for their hospitality. We thank A. Weiss and R. Bouwens for figures.

\clearpage
\newpage

\noindent  Amblard, A. et al. 2011, Nature, 470, 510

\noindent Andreon, S. \& Huertas-Company, M. 2011, A\& A, 526, 11

\noindent  Aravena, M., Bertoldi, F., Carilli, C. et al. 2010, ApJL, 
708, L36

\noindent  Baker, A. et al. 2004, ApJ, 613, L113

\noindent  Bauermeister, A. et al. 2010, 717, 323

\noindent  Blain, A. et al. 2002, PhR 369 111

\noindent  Bennett, C. et al. 1994, ApJ, 434, 587

\noindent  Bertoldi, F. et al. 2010  COSMOS mm

\noindent  Bigiel, F. et al. 2011, 730, L13

\noindent  Bournaud, F. et al. 2009, ApJ, 707, L1

\noindent  Bourne, N. et al. 2011, MNRAS, 410, 1155

\noindent  Bouwens, R. et al. 2010, ApJ, in press, arXiv 1006.4360

\noindent  Calzetti, D. et al. 1994, ApJ, 429, 582

\noindent  Carilli, C.L. et al. 2008, ApJ, 689, 883

\noindent  Carilli C.L. et al. 2010, 714, 1407

\noindent  Carilli C.L. et al. 2002, AJ, 123, 1838

\noindent  Carilli C.L. 2011, ApJ, 730, L30

\noindent  Carilli C.L. et al. 2011, ApJL, in press

\noindent  Carilli, C.L. et al. 2005, ApJ, 618, 586

\noindent  Chapman, S., Blain, A., Ivison, R., Smail, I. 2003 Nature
  422 695

\noindent  Chapman, S., Blain, A., Ibata, R. et al. 2009, ApJ, 691, 560

\noindent  Collins, D., Stott, J.P., Hilton, M. et al. 2009, Nature,
458, 603

\noindent  Coppin, K, et al. 2010, MNRAS, 407, L103

\noindent  Coppin, K, et al. 2010, ApJ, 665, 936

\noindent  Cormier, D. et al. 2010, A\& A, 518, L57

\noindent  Cowie, L. et al. 1997, 481, L9

\noindent  Daddi, E. et al. 2004, ApJ, 617, 746

\noindent  Daddi, E., Dickinson, M., Chary, R. et al 2005, ApJ 631 L13

\noindent  Daddi, E., Dannerbauer, H., Stern, D. et al. 
2009a, ApJ, 694, 1517

\noindent  Daddi, E., Dannerbauer, H., Krips, M. et al. 
2009b, ApJ, 695, L176

\noindent  Daddi, E. et al. 2010a, ApJ, 713, 686 

\noindent  Daddi, E. et al. 2010b, ApJ, 714, L118

\noindent  Dannerbauer, H., Walter, F., Morrison, G. 2008, ApJ, 673, L127

\noindent  Dave, R. et al. 2011, MNRAS, in press ArXiv1103.3528

\noindent  Dekel, A. et al. 2009, ApJ, 703, 785

\noindent  Downes \& Solomon 1998 ApJ, 507, 615

\noindent  Doherty, M., Tanaka, M., de Breuck, C. et al.
2009, A\& A, 509, 83

\noindent  Draine, B. 2003,   ARAA   41 , 241

\noindent  Dwek, E. et al.  2007,  ApJ   662 , 927

\noindent  Elvis, M. et al. 1994, ApJS, 95, 1

\noindent  Elvis, M. et al. 2002,  ApJ   567 , L107

\noindent  Elmegreen, B. et al. 2009, 692, 12

\noindent  Fan, X., Carilli, C. Keating, B. 2006, ARAA, 44, 415

\noindent  F\"orster-Schreiber, N. et al. 2011, ApJ, in press, arXiv1104.0248

\noindent  Geach, J. et al.  2011, ApJ, 730, L19

\noindent  Gao, Y. \& Solomon, P. 2004,  ApJ   606 , 271

\noindent  Gebhardt, K. et al. 2000, ApJ, 543, L5

\noindent  Genzel, R. \& Cesarsky, C. 2000, ARAA, 38, 761

\noindent  Genzel, R. et al. 2011, ApJ, in press ArXiv 1011.5360  

\noindent  Genzel, R. et al. 2010, MNRAS, 407, 2091 

\noindent  Genzel, R. et al. 2008, ApJ, 687, 59 

\noindent  Grazian, A. et al. 2007, A\& A, 465, 393

\noindent  Gultekin, K. et al. 2009, 706, 404

\noindent  Hainline, :. et al. 2006, ApJ, 650, 614

\noindent  H\"aring, N. \& Rix, W. 2004,  ApJ   604 , L89

\noindent  Harris, A. et al. 2010, ApJ, 723, 1139

\noindent  Ivison, R. et al. 2011, MNRAS,  412, 1913

\noindent  Kennicutt, R. 1998,  ARAA, 36, 189

\noindent  Keres, D., Yun, M.S., Young, J. 2003, ApJ, 582, 659

\noindent  Keres, D. et al. 2009, MNRAS, 395, 160

\noindent  Kormendy, J. \& Bender, R. 2011, Nature, 469, 377

\noindent  Krumholz et al. 2009, ApJ, 699, 850

\noindent  Kurk, J., Cimatti, A., Zamorani, G. et al. 2009,
A\& A, 504, 331

\noindent  Leroy, A. et al. 2008, 136, 2782

\noindent  Leipski, C. et al. 2010, A\& A, 518, L34

\noindent  Madau, P. et al. 1996, MNRAS, 283, 1388

\noindent  Maiolino, R. et al. 2005,   A\& A   40 , L51

\noindent  Malhotra, S. et al. 2001, ApJ, 561, 766

\noindent Marchesini, D. et al. 2009, ApJ, 701, 1765

\noindent Magnelli, B. et al. 2011, A\& A, 528, 35

\noindent Michalowski, M.  et al. 2010, A\& A, 522, 15

\noindent  Michalowski, M.  et al. 2010, A\& A, 515, 67

\noindent  Miley, G. \& de Breuck, C. 2008, A\& ARv, 15, 67

\noindent Momjian, E., Carilli, C. \& Petric, A. 2005, AJ, 129, 1809

\noindent Momjian, E. et al. 2010, AJ, 139, 1622

\noindent Moresco, M. et al. 2010, A\& A, 524, 67

\noindent  Murphy, E. et al. 2011, ApJ, in press arXiv1102.3920

\noindent  Narayanan, D. et al. 2011, ApJ, in press arXiv 1104.4118

\noindent  Obreschkow, D. \& Rawlings, S. 2009, ApJ, 696, L129

\noindent Pannella, M. et al. 2009, ApJ, 698, 116

\noindent  Papadopoulos, P. et al. 2002, ApJ, 564, L9

\noindent  Papadopoulos, P. et al. 2010, ApJ, 711, 757

\noindent  Perley, R. et al. 2011, ApJL, in press

\noindent Perley, D. et al.   ApJ  in press (2010)

\noindent  Renzini, A. 2006, ARAA, 44, 141

\noindent  Richards, G.T. et al. 2006, ApJS, 166, 470

\noindent  Riechers, D. et al. 2008, ApJ, 686, L9

\noindent  Riechers, D. et al. 2010a, ApJL, 720, L131

\noindent  Riechers, D. et al. 2010b, ApJ, 724, L153

\noindent  Riechers, D. et al. 2011a, ApJ, 726, 50

\noindent  Riechers, D. et al. 2011b, ApJ, 725, 1032

\noindent  Riechers, D. et al. 2011b, ApJ, in press arXiv1104.4348

\noindent  Sanders, D. et al. 1988, ApJ, ApJ 325, 74

\noindent  Schinnerer, E. et al. 2008, ApJ, 689, L5

\noindent  Scott, K. et al. 2011, ApJ, in press, arXiv1104.4115

\noindent  Shapiro, K. et al. 2008, ApJ, 682, 231

\noindent  Shapley, A. et al. 2003, ApJ, 651, 688

\noindent  Spitzer, L. 1998, {\sl Physical Processes in the
Interstellar Medium,} (Wiley)

\noindent  Solomon, P. \& Vanden Bout, P. 2005, ARAA, 43, 677

\noindent  Solomon, P. et al. 2003, Nature, 426, 636

\noindent  Stacey, G. et al. 2010, ApJ, 724, 957

\noindent  Stratta, G. et al. 2007,  ApJ   661 , L9

\noindent  Steidel, C. et al. 2004, ApJ, 604, 534

\noindent  Tacconi, L., Neri, R., Chapman, S. et al. 2006, ApJ, 640, 228

\noindent  Tacconi, L., Genzel, R., Smail, I., et al. 2008, ApJ, 680, 246

\noindent  Tacconi, L. et al. 2010, Nature, 464, 781  

\noindent  Thompson, T. et al. 2005,   ApJ   630 , 167

\noindent  Venkatesan, A. et al.   ApJ   640 , 31 (2006)

\noindent  Wagg, J. et al. 2010, A\& A, 519, L1

\noindent  Wagg, J. et al. 2005, ApJ, 634, L13

\noindent  Walter, F. et al.  2009, Nature   457 , 699

\noindent  Walter, F et al.  2004, ApJ   615 , L17 

\noindent  Walter, F. et al.  2003, Nature   424 , 406

\noindent Wang, R. et al.  ApJ   714 , 699 (2010b)

\noindent Wang, R. et al.  ApJ   687 , 848 (2008)

\noindent  Weiss, A. et al. 2007, ASPC, 375, 25

\noindent  Wootten, A. \& Thompson, A. 2009, IEEEP, 97, 1463

\noindent  Wu, J. et al. 2005, ApJ, 635, L173

\noindent  Yun, M.S. et al. 2001,  ApJ   554 , 803

\noindent  Zheng, X. et al. 2007, ApJ, 661, L41

\end{document}